\documentclass{rstransa} 
\usepackage{comment}
\begin{document}

\title{Effect of the vibration profile on shallow granular systems}
\author{Patricio Cordero$^1$, Dino Risso$^2$ and Rodrigo Soto$^1$}
\date{etc etc}
\address{$^1$ Departamento de F\'isica,
Facultad de Ciencias F\'isicas y Matem\'aticas, 
Universidad de Chile, $^2$ Departamento de F\'isica,
Facultad de Ciencias, Universidad del Bio-Bio}
\subject{Granular matter}
\corres{Patricio Cordero\\
\email{pcordero@dfi.uchile.cl}}

\jname{rsta}
\Journal{Phil. Trans. R. Soc}

\begin{abstract} 

We describe the collective behavior   of a system of many inelastic
spherical particles inside a box which is being periodically vibrated.  The
box is shallow, with large horizontal dimensions, while the height is less
than two particle diameters.  The vibrations are not symmetric: the time the
box is moving up is, in general, different to the time it is moving down. 
The limit cycles of isolated grains are largely affected by the asymmetry of
the vibration mode, increasing the size in phase space of the chaotic
regions.  When many grains are placed in the box, the phase separation
between dense, solid-like regions, coexisting with fluid-like regions takes
place at smaller global densities for asymmetric vibration profiles. 
Besides, the order parameter of the transition takes larger values when
asymmetric forcing is used.  \end{abstract}

\maketitle

\section{Introduction}

Granular systems, characterized by the dissipative collisional dynamics of
macroscopic objects still defy our understanding.  When energy is injected
at a high enough rate to compensate for the dissipated energy dynamical
states are generated.  Granular media, in many aspects, resemble molecular
fluids.  However, the energy dissipated at every collision permanently keeps
those systems in a non-equilibrium state.   The origin of the non-equilibrium states is
that energy injected by one mechanism---for example collisions with the
walls---is dissipated by a different one, interparticle collisions.  A
necessary condition for equilibrium would be detailed balance which is not
present in our system.

It has been shown that varying the injection mechanism generates
non-equilibrium states with different properties.  For example, the shear
viscosity varies from case to case.  Compare the outcomes in
Ref.\cite{viscosities}, where the Inelastic Hard Sphere model is used in all
cases, but the driving mechanism is different.  Hence it is interesting to study the
different behaviors of the system when different energy injection mechanisms
are used.

A particular geometry to study granular systems that has gained interest is
the quasi two-dimensional (Q2D) configuration.  In this case grains are put
in a shallow box:  large horizontal dimensions and a rather small
vertical one.  The box is vertically vibrated and the spherical grains gain
energy colliding with the top and bottom plates.  This energy is partly
transfered to the horizontal directions through collisions.  

Using shallow boxes has many advantages.  In experiments, it allows the
dynamics of the granular layer to be fully visualized by imaging the system,
as all particles can be seen and recorded with a camera from the top.  This
makes it possible to accurately study the system in a microscopic way, as
particle's positions and velocities can be measured for most particles at
any time.  Also the collective behavior can be
captured~\cite{98olafsen,ccdhmrv08,shallowhoriz1,shallowhoriz2,Pacheco}. 
Implied by the energy injection mechanism, the horizontal kinetic energy of
the grains can be quite different from the vertical kinetic
energy~\cite{anisotropia1,anisotropia2}.

When the height of the box is less than twice the particle's diameter  we
have a particular and relevant case: no two particles can be on top of one
another.  Here, gravity does not play a direct role in the horizontal
dynamics and the indirect effects can be controlled and reduced by means of
increasing the dimensionless acceleration of the box.  More importantly,
gravity is perpendicular to the eventual directions of phase separation,
allowing to discard many known mechanisms.  Nevertheless, experiments and
simulations have shown that segregation can also take place between
grains that differ in size or mass \cite{anisotropia2,Explosiones}.

It has been observed that a particular phase separation takes place in 
shallow boxes: grains form solid-like regions surrounded by fluid-like
ones having high contrasts in density, local order and granular
temperature~\cite{98olafsen}.  This phase separation is driven by the
negative compressibility of the effective two dimensional
fluid~\cite{ccdhmrv08} and, depending on the height of the box, it can be
either of first or second order \cite{Castillo1,Castillo2}.

In the present article we study the role that the energy injection plays in
the phase separation in Q2D systems.  To do so, keeping the material
properties and geometry of the system unchanged, we vary the waveform
profile of the vibration.  That is, we change the pure sinusoidal profile
where the upward and downward phases are symmetric, to asymmetric profiles
where both phases take different times.  Under these conditions we first
analyze the dynamics of isolated grains looking for the limit cycles that
each one of them develops.  For the case of an isolated particle with no top
wall see \cite{mehta90,Szymanski,barroso09}.  Next we consider dense regimes where the
grains interact so that phase separation may take place.  It is found that
asymmetric vibration profiles increase the size of the chaotic regions in
phase space and also facilitate phase separation.  Our main tool are event
driven molecular dynamics simulations, making use of our own quite efficient
strategy~\cite{mm1a}. See also~\cite{gonza}.

\section{The Model}

\subsection{System setup}

The grains are inelastic hard spheres with rotational degrees of freedom. 
These particles are in a vertically vibrated {\em shallow box}, namely we
consider granular systems in a box with a height of the order of two
particle diameters and much larger horizontal dimensions.  Collisions are
characterized by normal and tangential restitution coefficients as well as
friction coefficients.

\begin{figure}[htb]
\begin{center}
\includegraphics[width=.5\columnwidth,angle=0]{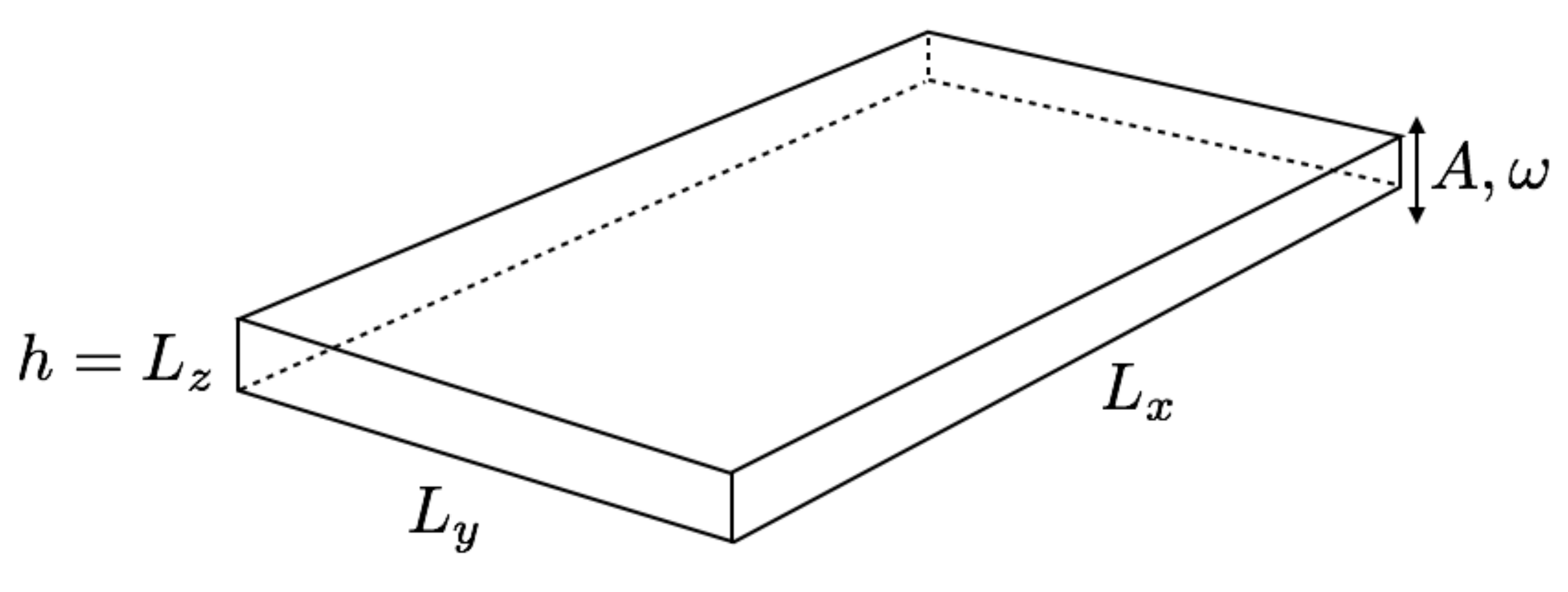}
\end{center}
\caption{Configuration of the quasi two-dimensional granular system. The box
is periodic in the horizontal directions, while it is confined by the top
and bottom hard walls. The inner height is $h$.  Grains are placed inside the box,
which is vertically  vibrated  with amplitude $A$ and angular frequency
$\omega$.}
\label{setup}
\end{figure}

The diameter $\sigma$ of the particles, the acceleration of gravity $g$ and
the mass $m$ of the particles are used to define the dimensionless
expressions in this article, except that time is measured in oscillation
periods, $T=2\pi/\omega$.  The horizontal lenghts of the box $L_x$ and $L_y$
will be different in different simulations, although in all cases they will
be large compared with the confinement height $h=L_z$, which is taken to be
between $1.8\sigma$ and $1.82\sigma$.  Simulations to study the phase
separations are done with $L_z\ll L_y\ll L_x$, where the system remains
homogeneous in the $y$ direction, allowing us to measure the density
contrast between the solid-like cluster and the surrounding liquid-like
phase.  The vibration is characterized by the amplitude $A$ and frequency
$\omega$.  Finally, collisions are characterized by the following mechanical
parameters: normal and tangential restitution coefficients $r_n$ and $r_t$,
while the static and dynamic friction coefficients are $\mu_s$ and $\mu_d$. 
Their specific values will be given in each case.
 
\subsection{Symmetric profiles: sinusoidal versus parabolic}

It may seem natural to consider sinusoidal vibrations, but we have studied
systems for which the vibration cycle is characterized by four separate
parabolic movements in the sense that the height depends quadratically in
time.  The acceleration is piecewise constant, with values $a=\pm
8A\omega^2/\pi^2$.  This parabolic movement is described in detail in the
next subsection.  It will be interesting to observe that for a given
vibration frequency there is a range of amplitudes for which the system
behaves almost as if there were sinusoidal vibrations.

\subsection{Asymmetric profiles}

In our studies energy is injected to the system by means of vertically
vibrating the box using asymmetric modes where, again the acceleration is
piecewise constant.  We define the {\em asymmetry coefficient} $\beta$,
$(0<\beta<1)$, which divides the period in two intervals of size $\beta T$
and $(1-\beta)T$.  In the first interval the box moves down with accelerations
$a=\pm 2A\omega^2/\beta^2\pi^2$, while in the second one the box moves up
with accelerations $a=\pm 2A\omega^2/(1-\beta)^2\pi^2$.

The height $Z$ of the top (+) and bottom (-) walls of the box are given by
\begin{equation}
   Z_\pm(t; \beta) = \pm \frac{h}{2} + \alpha (\tau;\beta)A,
\end{equation}
where $\tau=t{\rm\, mod\,} T$ and 
\begin{equation}\label{Alfa}
\alpha(\tau; \beta) = \left\{
\begin{array}{l c r}
\frac{1}{\beta^2 T^2}(\beta^2 T^2 -  4\tau^2) & \qquad & 0 \le \tau \le \frac{\beta T}{2}\\ & \\
   \frac{1}{\beta^2 T^2}\,(2\tau-\beta T)(2\tau-3\beta T)& & \frac{\beta T}{2} \le \tau \le \beta T \\ & \\
   \frac{1}{(1-\beta)^2\,T^2}\,(2\tau +(1-3\beta) T)(2\tau - (1+\beta)T) & & \beta T \le \tau \le \frac{(1+\beta)T}{2} \\ & \\
   \frac{1}{(1-\beta)^2\,T^2}\,(2\tau- (3-\beta)T  )((1+\beta)T-2\tau) & & \frac{(1+\beta)T}{2} \le \tau \le T
\end{array}
\right.
\end{equation}

This function is continuous and has continuous first derivatives.  The first
derivative of $\alpha(\tau; \beta)$ vanishes at $\tau=0$, $\tau=\beta T$ and at $\tau=T$. 
For $\beta=\frac{1}{2}$ this quadratic function roughly resembles $\cos(2\pi
\tau/T)$.

\begin{figure}[htb]
\begin{center}
\includegraphics[width=.8\columnwidth]{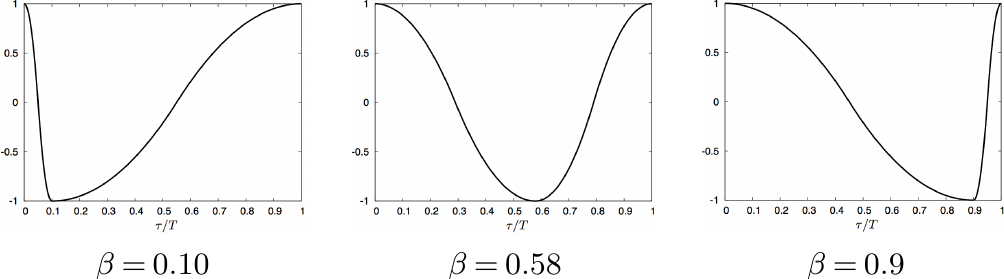}
\end{center}
\caption{The function $\alpha(\tau; \beta)$ that defines the
box-oscillation profile for different values of the asymmetry parameter
$\beta$.  For $\beta<1/2$ the downwards interval takes a shorter time than
the upward movement (left) and conversely for $\beta>1/2$ (right).  For
$\beta=1/2$  the function $\alpha(\tau; \beta)$ resembles
$\cos(2\pi\tau/T)$.
} \label{perfiles}
\end{figure}

The control parameters are the asymmetry parameter $\beta$, the amplitude
$A$ and angular frequency $\omega= 2\pi/T$ of the oscillations.  The latter
is characterized by the dimensionless acceleration $\Gamma = A\omega^2/g$
and dimensionless velocity $\zeta =A\omega/\sqrt{g\sigma}$.

\section{A single particle in the vibrating box}

In the case of shallow boxes it is important to understand first the
movement of a single particle.
There are conditions under which the particle ends up in permanent contact
with the base of the box.  But if  $A\omega^2/g$
is larger than unity such situation cannot occur.  Next, it may happen that
the energy injected to the particle by the basis is not enough for the
particle to reach the upper wall in the long term evolution.  We consider conditions under which
neither of these last two cases can take place.

If the single particle is left bouncing for a sufficiently large time, its
angular and horizontal translation velocities vanish so that the movement of
the particle becomes strictly one dimensional and the friction coefficients
as well as the tangential restitution coefficient become irrelevant.  There is a
wide range of amplitudes and frequencies for which the particle reaches this
quite simple limit cycle characterized by two velocities: the velocities
with which the particle bounces off the two horizontal walls.

\begin{figure}
\includegraphics[width=\columnwidth]{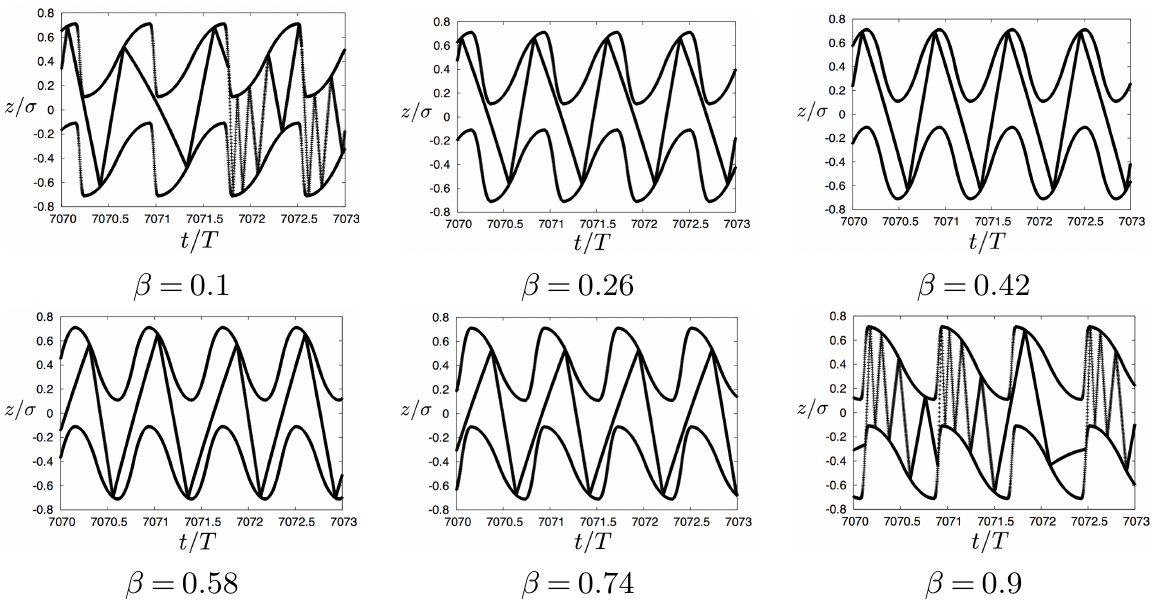}
\caption{A single particle in the shallow box subject
to vibrations having the same $\Gamma$ and $\zeta$ but different values of $\beta$
may have quite different stationary bouncing modes. In this example we chose
$\Gamma=18.56$ and $\zeta=2.32$.} 
\label{fig3}
\end{figure}

\begin{figure}
\begin{center}\includegraphics[width=.9\columnwidth]{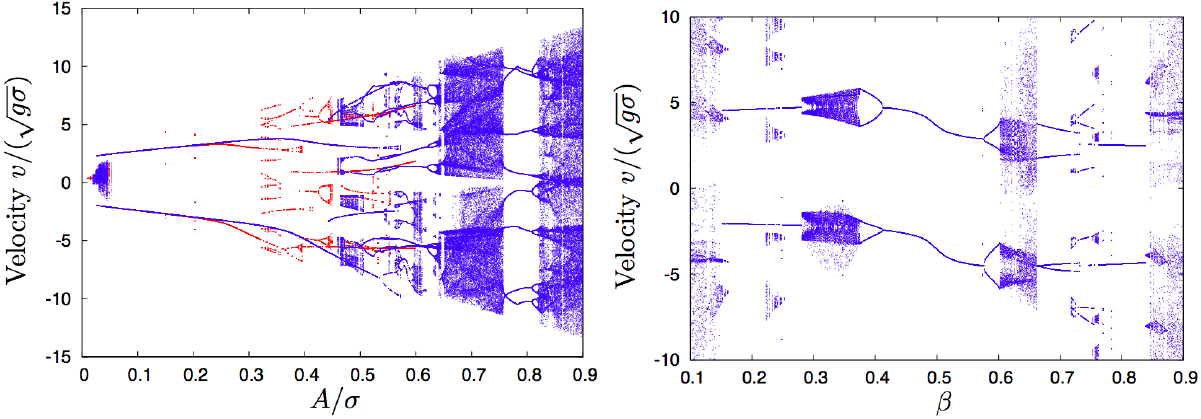}
\end{center}
\caption{Bouncing velocities after colliding with the top and bottom walls
of the shallow box once the particle has reached a steady state.  Left:
Varying the vibration amplitude $A$ in the case of the sinusoidal (red) and
bi-parabolic (blue) symmetric profiles ($\beta=1/2$).  Right: Varying the
asymmetry parameter $\beta$ with fixed vibrating amplitude $A/\sigma=0.3$. 
The asymmetry on the obtained velocities around $\beta=1/2$ is due to the
presence of gravity.  In both figures, $\omega=8\sqrt{g/\sigma}$, $r_n=0.8$,
and $h=1.82\sigma$.  In the right figure $\Gamma=19.2$ and $\zeta=2.4$. } \label{fig2}
\end{figure}

Increasing the vibration frequency or the amplitude the limit cycle of the
single particle reaches a threshold above which the limit cycle suffers a
bifurcation.  Now the grain bounces twice at the top and bottom walls before
repeating its movement.  Hence the new limit cycle takes twice the time
associated to the limit cycle just before the bifurcation.  There is a chain
of bifurcations of this type until the bouncing of the single particle
becomes chaotic.  Some cases of the resulting limit cycles are presented in
Fig.\ \ref{fig3}.  All this can be appreciated in Figure \ref{fig2} where it
is seen that other modes appear as the amplitude grows, that is, the
particle has more than one bouncing mode for a given value of the amplitude. 
Figure \ref{fig2} includes in red the bouncing velocities when the vibration
is sinusoidal with the same frequency and amplitude as our parabolic
vibration.  To our surprise, when $\beta=0.5$ the bouncing velocities are
almost exactly the same till about $A/\sigma=0.2$.

\begin{figure} \begin{center} \begin{tabular}{ccc}
\includegraphics[width=\columnwidth]{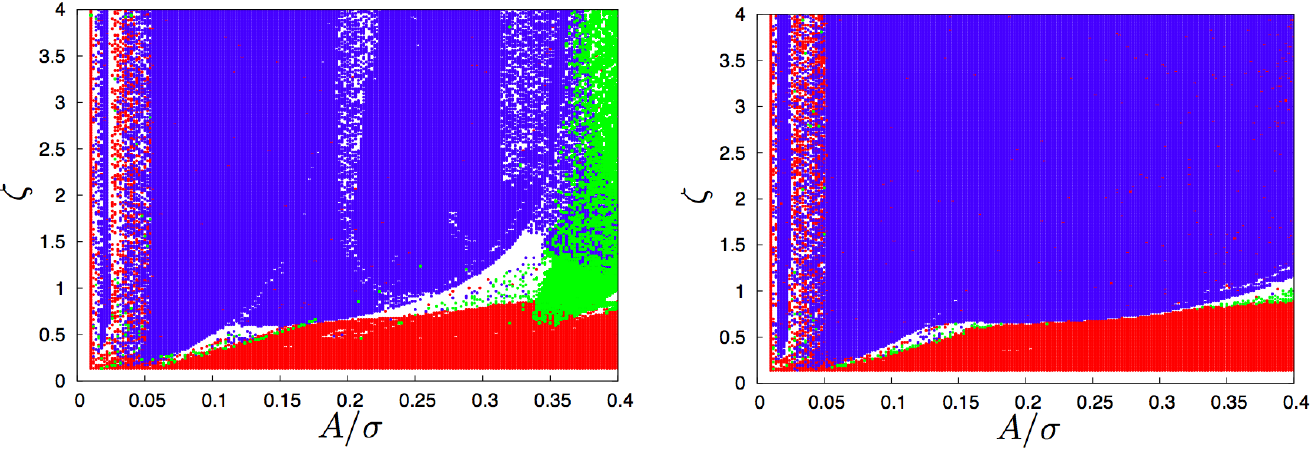}
\end{tabular} \end{center} \caption{ Map in the $A$-$\zeta$ plane of the
qualitative characterization of the limit cycles for the sinusoidal profile
(left) and for the symmetric $\beta=1/2$ parabolic profile (right).  The
possible outcomes are: bouncing only with the bottom wall with period $T$
(red), bouncing with the top and bottom walls with period $T$ (blue),
bouncing with the bottom and top walls with double period $2T$ (green), and
more complex states (white).  In both cases $h=1.8\sigma$ and $r_n=0.8$.  }
\label{fig2b} \end{figure}

Figure \ref{fig2b} presents a qualitative picture of the possible outcomes. 
Different colors are used to represent the single period, double period, and
chaotic motion cases.  The comparison shows how similar is the behavior with
sinusoidal and parabolic profiles, except that for large amplitudes the
parabolic profiles are more regular.  In both cases, a band of chaotic
motion is present for small amplitudes.

\begin{figure}
\begin{center}
\includegraphics[width=\columnwidth]{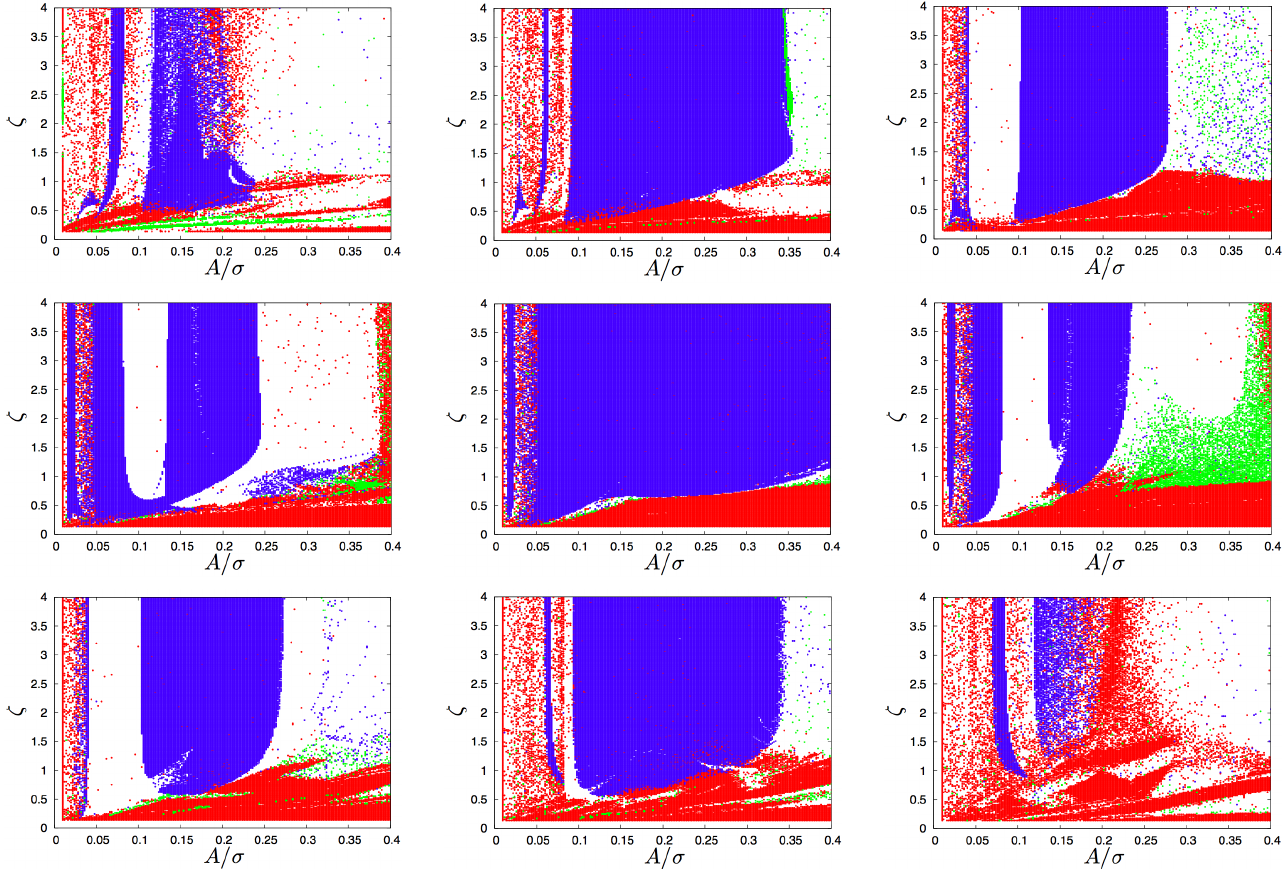}
\end{center}
\caption{Map in the $A$-$\zeta$ plane of the qualitative characterization of
the limit cycles for different values of $\beta$ using the same color coding
as in Fig.\ \ref{fig2b}.  From left to right, top: $\beta=0.1, 0.2, 0.3$,
middle: $\beta=0.4, 0.5, 0.6$, bottom: $\beta=0.7, 0.8, 0.9$).  In
all cases $h=1.8\sigma$ and $r_n=0.8$.
}
\label{mapa-ptos-fijos}
\end{figure}

Using the same analysis, we consider the asymmetric profile, varying the
parameter $\beta$ in the case where in the symmetric profile the limit cycle
consists on a single bounce on the top and bottom walls.  Figure \ref{fig2}
shows that close to the symmetric case ($\beta\approx 0.5$) the limit cycle
is perturbed, keeping its qualitative behavior but changing the values of
the bouncing velocities.  Further away from the symmetric condition, period
doubling and chaos appears.  The qualitative plots in Figure
\ref{mapa-ptos-fijos} confirm this result. 
The most regular
motion is obtain in the vicinity of $\beta=0.5$.
For intermediate values of $\beta$ figure
\ref{mapa-ptos-fijos} shows that the effect of gravity is obvoius in the
case of smaller values of $\zeta$ while for $\zeta$ large (small $g$) 
the structure of the figure is weekly dependent of $\zeta$.

\section{Phase separation}

\begin{figure}[htb]
\begin{center} 
\includegraphics[width=.8\columnwidth]{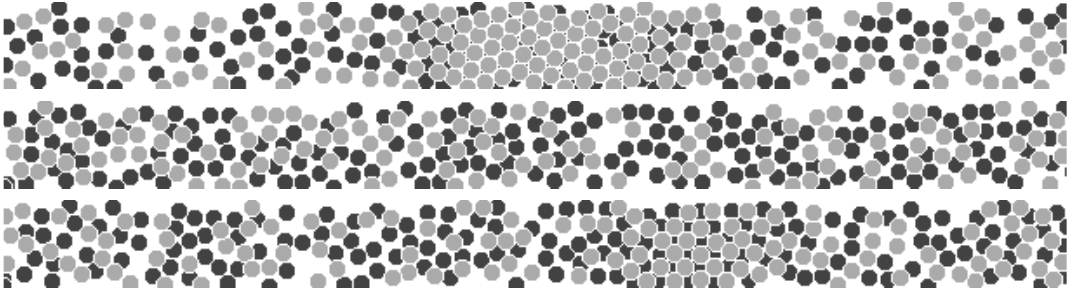}
\end{center} \caption{Top view of a shallow system in a rather
narrow box ($L_y\ll L_x$) in a typical state for a system with common values
for all their parameters except for the value of $\beta$.  Particles are all
the same, but they are represented by black circles if their height is
bellow half the height of the box, otherwise they are represented by gray
circles.  From top down $\beta=0.1$, $\beta=0.5$ and $\beta=0.75$.  The
first and third cases have notoriously stable phase separated states,
showing large density contrasts.  In all cases $N=410$,
    $L_x=90.01\sigma$,
    $L_y=4.65\sigma$,
    $h=1.8\sigma$,
    $r_n=r_t=0.91$, 
     $\mu_d=0.10$,
    $\mu_s=0.17$,
    $\omega=10\sqrt{g/\sigma}$,
    and 
    $A=0.29\sigma$ implying $\zeta=2.9$ and $\Gamma=29.0$. 
}
\label{quasi1D}
\end{figure}

In the case of systems composed by a single species of granular matter there
are conditions under which the system separates into high and low density
zones \cite{ccdhmrv08}.  The results of the previous section illustrate a
central point: the form of the vibration---determined by the parameter
$\beta$---has an important effect in the way the system behaves.  This is
illustrated in Figure \ref{quasi1D} which shows a characteristic state of a
vibrated granular system in a narrow 3D box: depending on the value of
$\beta$ the system may or may not present phase separation.  Indeed, with
all other parameters fixed to a case where the symmetric case would yield to
a state with no phase separation, forcing with asymmetric profiles induces
phase separation.

\begin{figure}[htb]
\begin{center}
\includegraphics[height=0.45\textwidth]{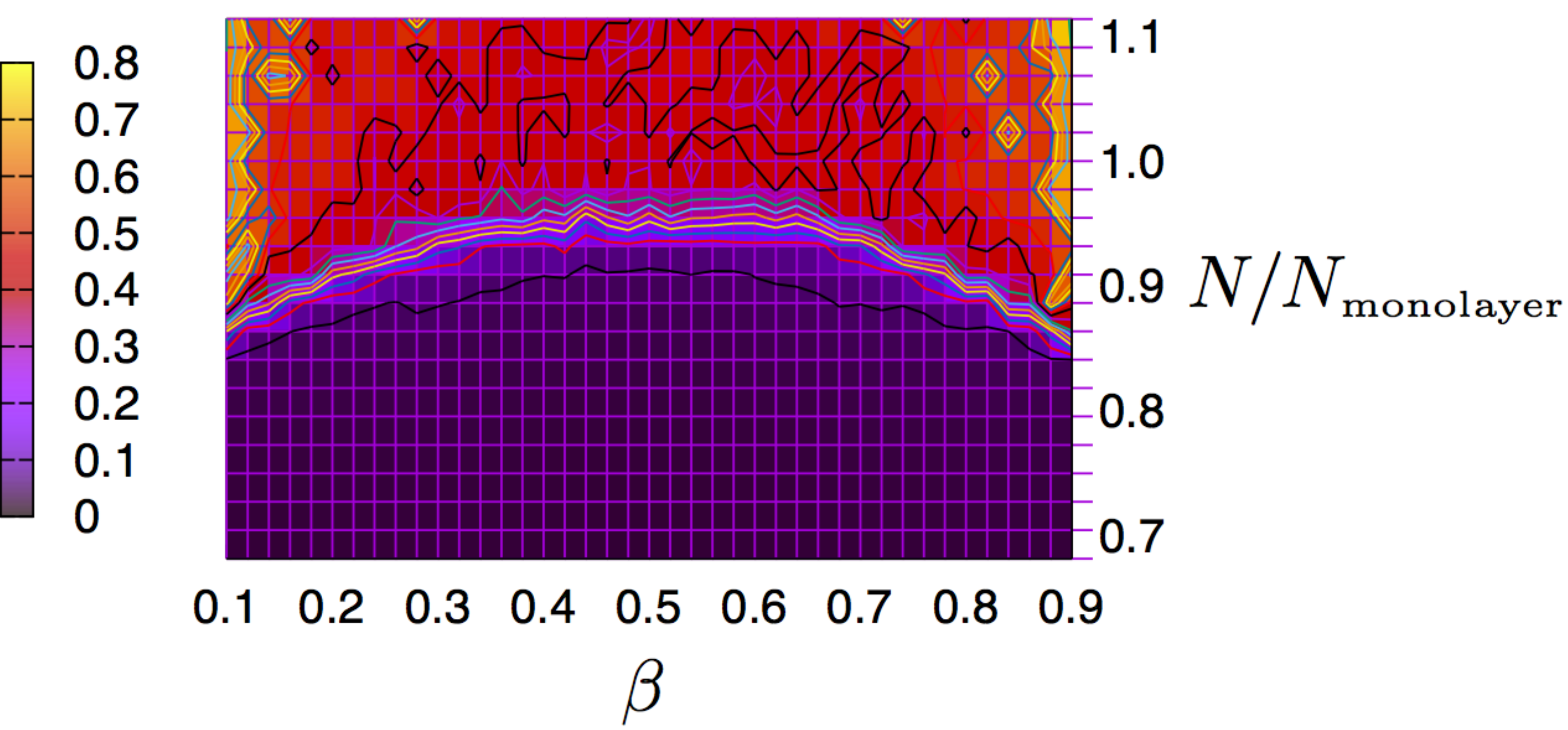}
\end{center}
\caption{
Density contrast $c$ in the $\beta-N/N_{\mbox{\tiny monolayer}}$ plane for 
a system with $L_z=1.8\sigma$ and  $L_x\times L_y=90\sigma\times 4.65\sigma$
and vibration parameters $\tilde{\omega}=\omega/\sqrt{g/\sigma}=3$  and
$A=0.29\sigma$; $\zeta=0.87$,  $\Gamma=2.61$.  
The inelasticity and friction grain-grain and grain-wall coefficients are 
$r_n=r_t=0.91$ and $\mu_d=0.10$, $\mu_s=0.17$. 
}
\label{contrast1}
\end{figure}

For a quantitative analysis, we define the {\em contrast}  observable $c$,
which is the ratio between the smaller and larger observed area densities:
$c=(n_{\mbox{\tiny max}}-n_{\mbox{\tiny min}})/(n_{\mbox{\tiny
max}}+n_{\mbox{\tiny min}})$.  In the case of the system shown in Fig.\
\ref{quasi1D}, which are strictly 3D, the density is evaluated considering
many transversal strips and calculating the area density in each strip,
defined using the projection of the particles in each 2D strip.

\begin{figure}[htb]
\includegraphics[width=\columnwidth]{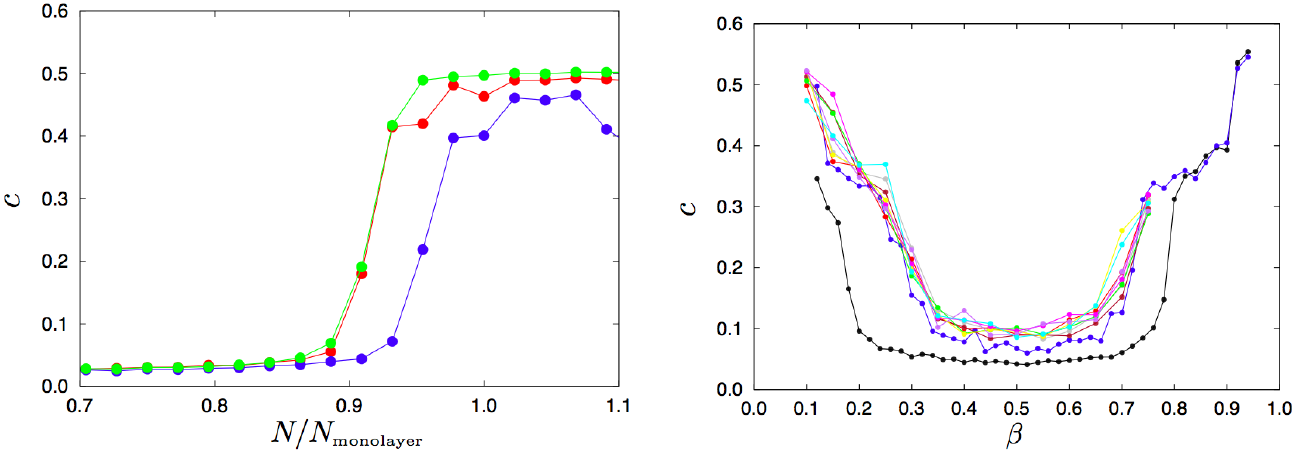}
\caption{
Left: density contrast against $N/N_{\mbox{\tiny monolayer}}$ for
$\beta=0.22$ (red), $\beta=0.5$ (blue) and $\beta=0.78$ (green).  The
amplitude is fixed to $A=0.29\sigma$ while the dimensionless frequency takes
the value $\tilde{\omega}=\omega/\sqrt{g/\sigma}=3$.  Right: density
contrast against $\beta$ for a system with $N=410$ grains ($N/N_{\mbox{\tiny
monolayer}}=0.93$).  The amplitude is fixed to $A=0.29\sigma$ while the
dimensionless frequency takes the values
 $\tilde{\omega}=2$ (black), 
 $\tilde{\omega}=3$ (blue), 
 $\tilde{\omega}=5$ (grey), 
 $\tilde{\omega}=10$ (red), 
 $\tilde{\omega}=15$ (brown), 
 $\tilde{\omega}=20$ (green), 
 $\tilde{\omega}=25$ (yellow),
 $\tilde{\omega}=30$ (magenta), 
 $\tilde{\omega}=35$ (purple) 
 and
 $\tilde{\omega}=40$ (cyan).  
In all cases $L_z=1.8\sigma$, $L_x\times
L_y=90\sigma\times 4.65\sigma$ and the inelasticity and friction grain-grain
and grain-wall coefficients are $r_n=r_t=0.91$ and $\mu_d=0.10$,
$\mu_s=0.17$.
}\label{contrast2}
\end{figure}

Figure \ref{contrast1} shows that, in agreement with the snapshots in Fig.\
\ref{quasi1D}, the density contrast is larger when $\beta$ is away from the
symmetric case $\beta=0.5$.  A transition line can be identified where there
is steep increase in $c$.  This transition line indicates that away from the
symmetric case, a smaller number of particles is needed to produce the phase
separation.  This is clearly presented in the left panel of Fig.\
\ref{contrast2}, where the contrast is shown as a function of the number of
particles for three values of $\beta$.  A clear transition is observed, with
a critical number of particles that is larger for $\beta=1/2$.

Figure \ref{contrast2}-right presents the contrast $c$  as a function of
$\beta$ for a fixed number of particles and different values of $\omega$. 
The collapse between the majority of the curves shows that the mayor impact in the phase
separation is the shape of the vibration rather than the energy injection
rate.

\section{Conclusions}

In this article we have studied the role of changing the energy injection
profile of the box in vibrofluidized shallow granular media, where the box
moves vertically in a periodic way.  With this aim we modify the oscillation
profile.  Originally, the profile was symmetric with piecewise constant
accelerations, mimicking a sinusoidal oscillation.  This profile was changed
to an asymmetric one, still with piecewise constant accelerations, where the
downward and upward phases take different fractions of the oscillation
period: $\beta$ and $1-\beta$, respectively.

Single isolated grains reach states where the motion is purely vertical. 
These states can be periodic in the form of limit cycles with the same
period as the box or with a period that is an integer multiple of the box
period.  Also chaotic or irregular motion is possible.  The effect of
moving toward asymmetric forcing is to increase the region in parameter
space where chaotic motion is observed.  Conversely, it is more probable to
find regular motion with the same period as the box for the symmetric
forcing.

\begin{figure}[htb]
\begin{center}
\includegraphics[angle=0,width=0.4\textwidth]{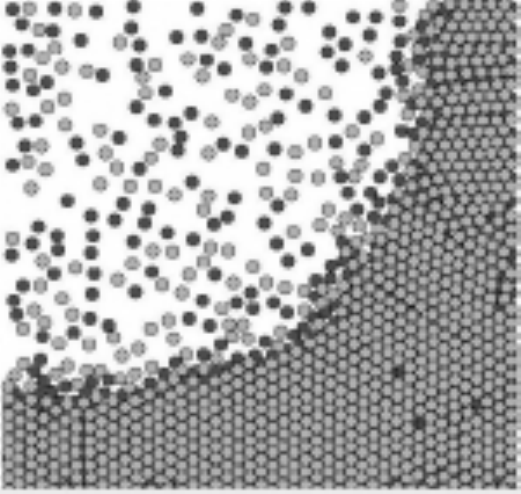}
\end{center}
\caption{Top view of a  shallow system in a box of square basis.  This is a
typical state once phase separation has taken place.  The black/grey
coloring is as in Fig.\ \ref{quasi1D}.  The limit of infinitely rapid
vibration is taken ($A\to0$, $\omega\to\infty$, while $A\omega=V_0={\rm
constant}$) for a sawtooth profile with no gravity.  This state was
obtained using $N=1732$, $L_x=L_y=40\sigma$, $L_z=1.9\sigma$, $r_n=r_t=0.7$
(grain-grain), $r_n=r_t=0.999$ (wall-grain), and $\mu_s=\mu_d=0.1$.}
 \label{qaasi2D}
\end{figure}

The asymmetric forcing also has an effect on the phase separation between
dense and dilute regions when many particles are placed in the box.  The
phase separation takes place for smaller number of particles when the
forcing is asymmetric.  The previous result suggests that other ways of
increasing the asymmetry could also be efficient in producing phase
separation.  As an extreme case, we consider a sawtooth profile at high
frequency and small amplitude, such that particles always meet the walls at
the same position and approaching the particles with a fixed velocity. 
Figure \ref{qaasi2D} shows that indeed phase separation takes place and 
high contrast states can be attained.

It remains as an open question if there is a relation between the accessibility to chaotic
behavior for single particles far from the symmetric profile and the
enhanced phase separation in the collective dynamics with many particles.

In Ref.\ \cite{ccdhmrv08} it was shown that the origin of the phase
separation is a negative compressibility region in the effective equation of
state of the pressure as a function of density $p=p(n)$.  This equation of
state emerges as a consequence of the temperature being enslaved to the
density under high vertical confinement, with $T(n)$ a monotonically
decreasing function.  Then, equations of state $p=p(n,T)$ derived from
kinetic theory, with positive compressibility can turn into effective
equations of state with negative compressibility and van der Waals loops
\cite{vdW2}.  An attempt to understand why is it that for $\beta$ far from
$0.5$ a phase separation takes place for lower densities it would be
necessary to study in the stationary regime the enslaving $T(n)$, which will
depend on $\beta$.  This analysis needs the study of the limit cycles
coupled with the three-dimensional collisions that transfer energy from the
vertical to the horizontal degrees of freedom.  This analysis, beyond the
scope of this article, is postponed for a further study

The result of the present article as well as others confirm that the
properties of non-equilibrium systems depend not only on the internal
dynamics but also on the energy injection mechanism.  Consistent
with this it should be said that there is no guarantee that the conclusions
raised here could be valid for other forcing mechanisms.

\section{Authors contributions}

P. Cordero and D. Risso carried out the simulations. All authors
participated in the design of the simulations, analysis the results and
writing of the manuscript.

\funding{This research was supported by Fondecyt Grants Nos. 1120775 and 1140778.}

\end{document}